\documentstyle[sprocl]{article}

\input{epsf}
\begin{document}

\title{MAGNETOTRANSPORT IN QUASILATTICES}

\author{\sc Uwe Grimm, Florian Gagel,  Michael Schreiber}

\address{Institut f\"ur Physik, Technische Universit\"{a}t, 
D-09107 Chemnitz, Germany}

\maketitle\abstracts{ The dc conductance and the Hall voltage of
planar arrays of interconnected quantum wires are calculated
numerically.  Our systems are derived from finite patches of aperiodic
graphs, with completely symmetric scatterers placed on their vertices
which are interconnected by ideal quantum wires.  Already in a
periodic square lattice arrangement, quantum interference effects lead
to complicated magnetotransport properties related to the Hofstadter
butterfly. For rectangular Fibonacci grids and other quasiperiodic
lattices, we obtain still more complex fractal patterns.  In
particular, irrational ratios of edge lengths and of tile areas in our
samples destroy the periodicities with respect to the Fermi wave
vector and the magnetic flux, respectively.  }

\section{Introduction}

Phase-coherent quantum transport manifests itself through many
exciting phenomena. The observation of conductance quantization, the
Aharonov-Bohm effect, universal conductance fluctuations, and other
interference effects in mesoscopic systems (see, e.g., Ref.~$1$ for an
overview) has been made possible by the technological advance in
device fabrication.  In this domain, macroscopic properties like
conductance and Hall voltage critically depend on the interference of
multiply scattered waves. Principally, this opens up the possibility
to extract information about a system through {\em conductance
spectroscopy}, i.e., by observing its transport properties upon
variation of the microscopic interference pattern, for instance via an
external magnetic field.

Even in a simple periodic structure, the presence of a magnetic field
leads to complicated fractal spectra. A well-known example is the
famous Hofstadter butterfly$^{\, 2}$ which describes the eigenvalue
spectrum of electrons in a periodic lattice subject to a perpendicular
magnetic field.  Here, we investigate phase-coherent magnetotransport
in quasiperiodic structures. As model systems, we choose Fibonacci
grids and finite quadratic patches of the ideal octagonal
Ammann-Beenker tiling (compare Fig.~1).  While related work is mostly
based on a tight-binding approach,$^{3-5}$ we follow Ref.~$6$ and
consider an ensemble of interconnected scatterers that are linked by
ideal waveguides.  The advantage is that we do not depend on the
details of the leads that have to be specified for any open
tight-binding system.  Our approach can be seen as a ``strong
confinement'' limit, where one-dimensional electron motion is possible
only on straight, ideal connection lines between the crossings.

\section{Numerical approach}  
 
\begin{figure}[tb]
\centerline{\epsfxsize=0.45\textwidth\epsfbox{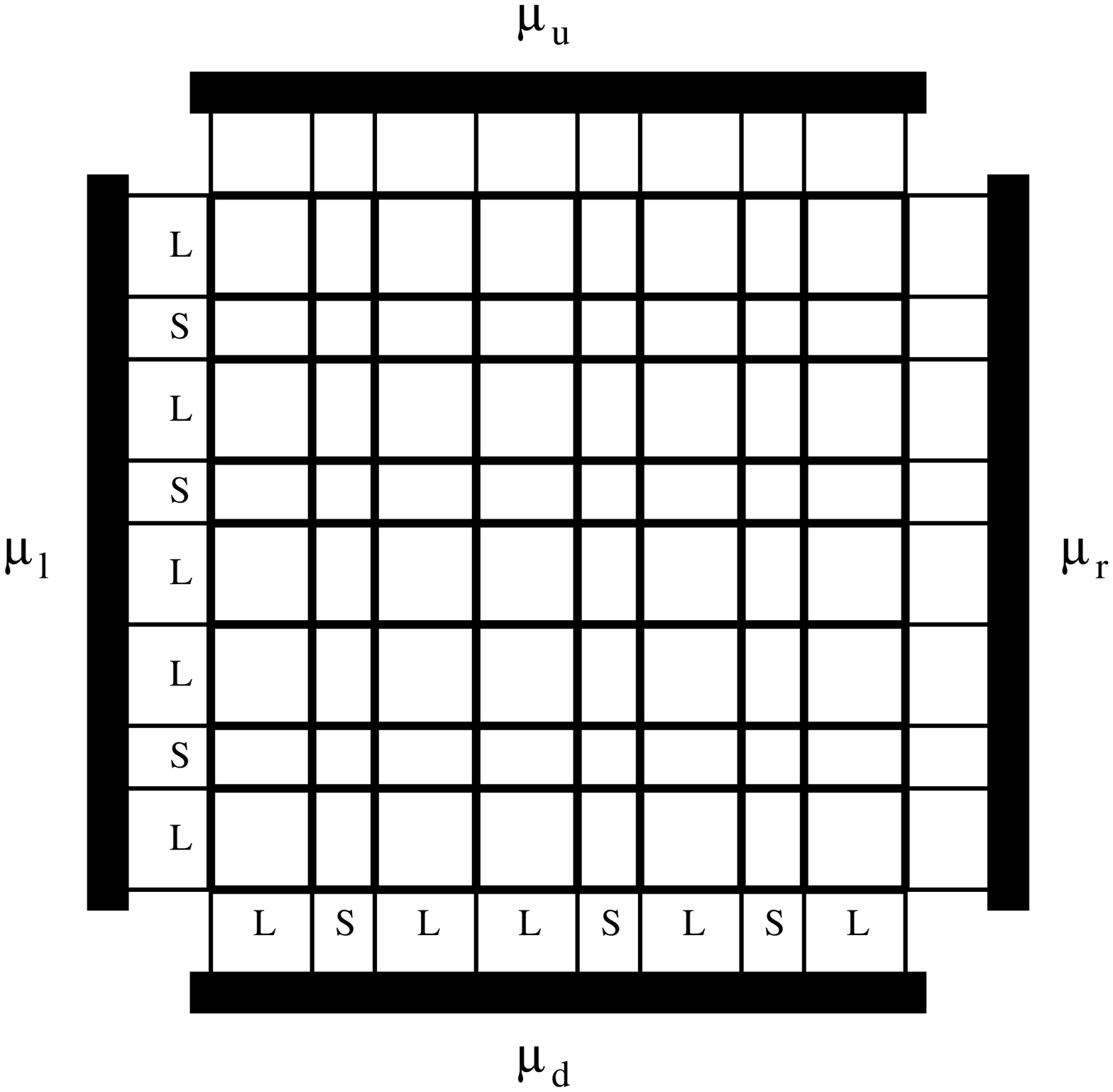}
\hfill
\epsfxsize=0.45\textwidth\epsfbox{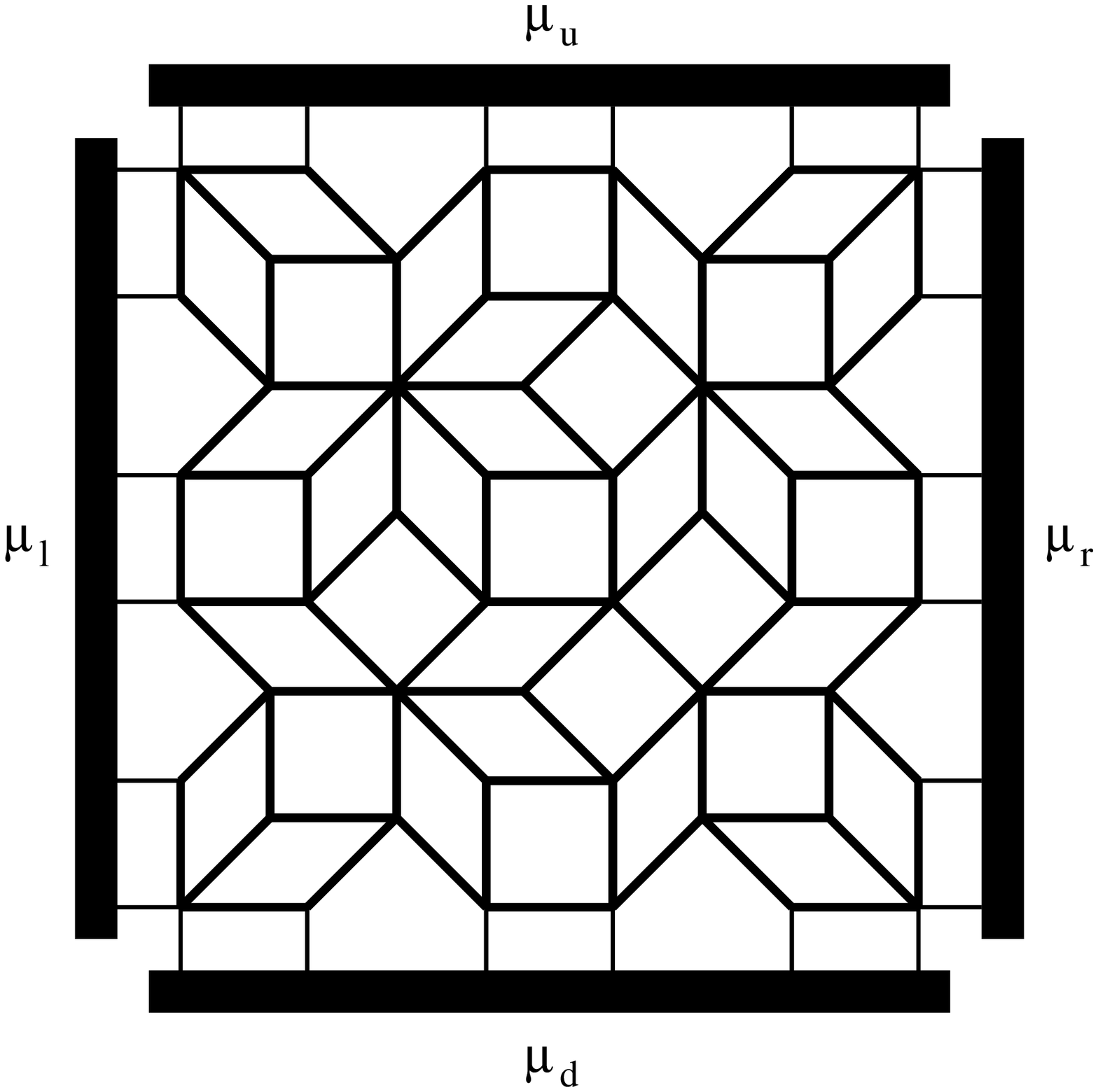}}\vspace*{-1ex}
\caption{Geometry of our systems:
Fibonacci grid (left) and octagonal approximant (right).\label{fig1}}
\end{figure}

We consider quadratic systems connected to four electron reservoirs
(terminals), see Fig.~\ref{fig1}.  Assuming small driving voltages
(applied between the left and right terminals, the ``up'' and ``down''
contacts serving as voltage probes), we have a linear relation between
currents and chemical potentials of the contacts,
\begin{equation} \label{eq:1} 
  I_\alpha^{}\; =\; \frac{2 e}{h}\;\sum_{\beta}\: 
T_{\alpha\beta}^{} \, \mu_\beta^{}
\qquad\quad
\mbox{($\alpha,\beta\in\{\mbox{l},\mbox{r},\mbox{u},\mbox{d}\}$)}
\end{equation}
where $I_\alpha^{}$ is the outgoing current at terminal $\alpha$ and
$\mu_\beta^{}$ denotes the chemical potential of terminal $\beta$.
This system of equations has to be solved while ensuring current
conservation in the voltage probes.  The voltage between two contacts
is related to the difference in their chemical potentials by
$eU_{\alpha\beta}^{} = \mu_\beta^{}-\mu_\alpha^{}$.

In the Landauer-B\"uttiker description,$^{7}$ the transmission matrix
$T$ is obtained from the {\em scattering matrix\/} $S$.  First, we
consider a node connecting $N$ waveguides.  In order to describe the
scattering process at such a crossing, we use a symmetric $N\times N$
scattering matrix $S_N^{}$ with diagonal elements (backscattering
amplitudes) $r$ and off-diagonal elements (transmission amplitudes)
$t$.  Neglecting spatial extension, assuming time-reversal symmetry
and demanding unitarity implies $t=2/N$ and $r=t-1$.  In the limit of
a continuum of scattering channels, $S_N$ corresponds to isotropic
$S$-wave scattering.

Now, we consider a pair of adjacent linked nodes $j$ and $k$. A wave
outgoing from node $k$ acquires a phase factor $P_{jk}^{}=\exp[i(q
a_{jk}^{}+\varphi_{jk}^{})]$ along the waveguide between the two
nodes. Here, $q$ denotes the modulus of the wave vector, and
$a_{jk}^{}=a_{kj}^{}$ measures the distance between the two nodes.
The contribution $\varphi_{jk}^{}=\frac{e}{\hbar}\int_k^j \vec{A}\cdot
d\vec{s}$ accounts for the additional phase shift due to the magnetic
field, and the Landau gauge can be used for the vector potential:
$\vec{A} = B y \vec{e}_x^{}$.

\begin{figure}
\centerline{\epsfxsize=0.45\textwidth\epsfbox{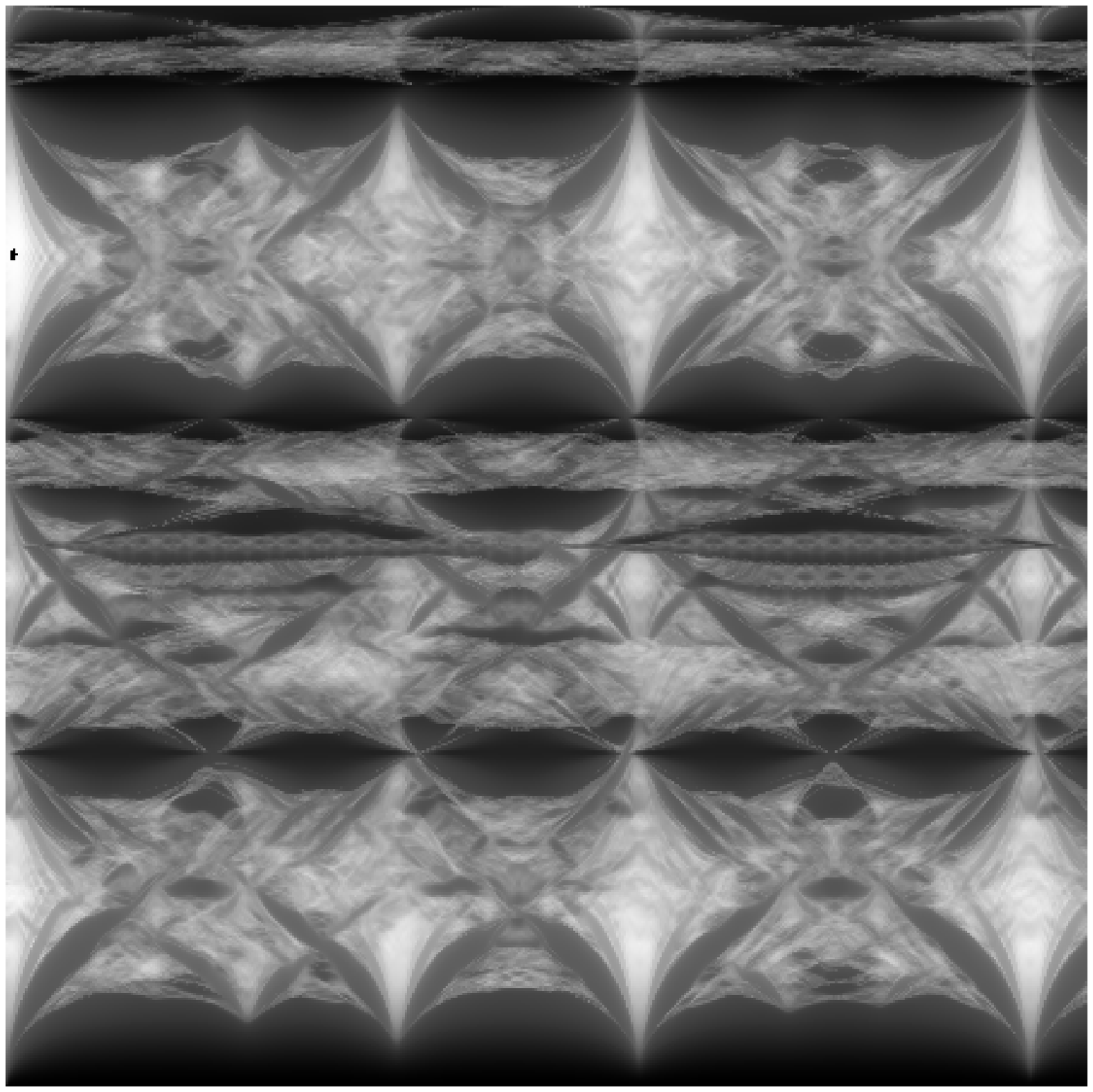}
            \hfill
            \epsfxsize=0.45\textwidth\epsfbox{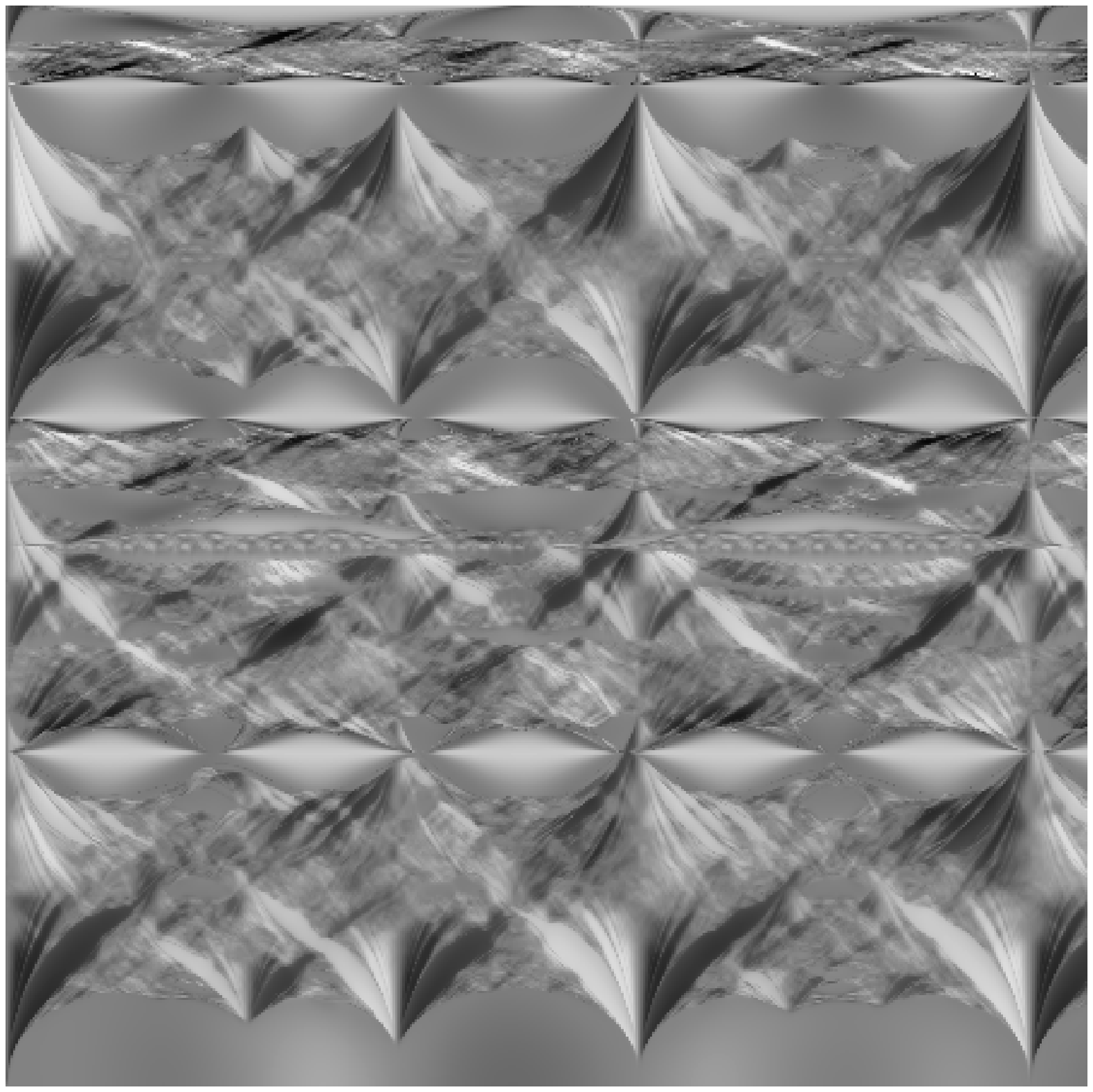}}\vspace*{-1ex}
\caption{Conductance (left) and Hall voltage (right) for a Fibonacci grid
as a function of the wave vector $q$ (vertically from $0$ to $2\pi$)
and the magnetic flux $\Phi$ 
in units of the flux quantum $h/e$ 
(horizontally from $0$ to $2$).
On the left side, the conductance vanishes in the dark regions and increases
with the brightness. In the right plot, the Hall voltage is negative 
in the dark parts and positive in the bright 
regions.\label{fig2}\hspace*{\fill}}
\end {figure}

In order to obtain the {\it overall\/} scattering matrix $S$ from the
$S_N^{}$ and $P_{jk}^{}$, we use the approach presented in Ref.~$6$ to
sum up all the multiple interferences in the system.  Then we solve
Eq.~(\ref{eq:1}) under the condition of current conservation in the
two voltage contacts ($I_{\mbox{\scriptsize
u}}^{}=I_{\mbox{\scriptsize d}}^{}=0$).  This yields the
magnetoconductance $G = I/U$, $I=I_{\mbox{\scriptsize
l}}^{}=-I_{\mbox{\scriptsize r}}^{}$ being the total current and
$U=U_{\mbox{\scriptsize rl}}^{}$ the voltage measured between source
and drain contact, as well as the Hall conductance $G_H^{} =
I/U_H^{}$, where $U_H^{}=U_{\mbox{\scriptsize ud}}^{}$ is measured
between the two voltage probes.

\section{Results}

\begin{figure}
\centerline{\epsfxsize=\textwidth\epsfbox{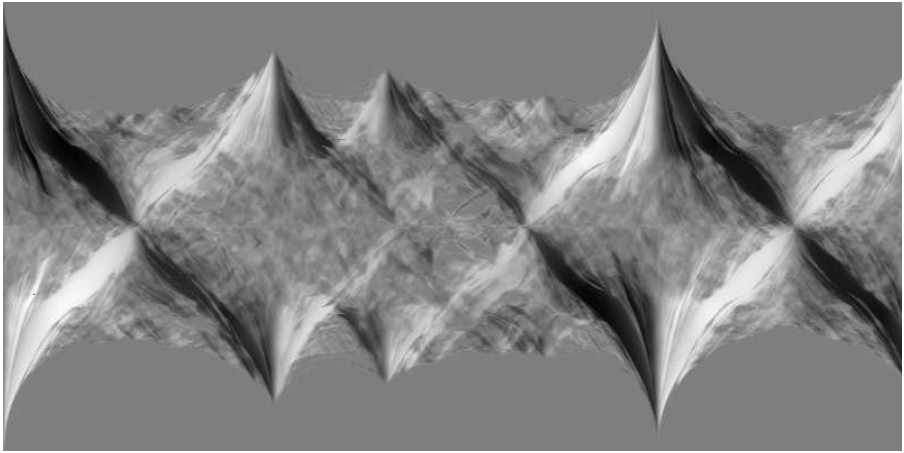}}\vspace*{-1ex}
\caption{Hall voltage for an octagonal patch
as a function of the wave vector $q$ (vertically from  $0$ to $\pi$)
and the magnetic flux (horizontally from $0$ to $4$).
The shading is as indicated in Fig.~\ref{fig2}. 
Note the almost perfect antisymmetry
with respect to $q=\pi/2$.\label{fig3}\hspace*{\fill}}
\end {figure}

In Fig.~\ref{fig2}, the conductance and the Hall voltage for a
Fibonacci grid with $14^2=196$ scatterers is shown as a function of
the wave vector and the magnetic flux. The Fibonacci grid contains
edges of two different lengths [$1$ and the golden mean
$\tau=(1+\sqrt{5})/2$] and tiles of three different areas ($1$, $\tau$
and $\tau^2=\tau+1$).  Fig.~\ref{fig3} shows the Hall voltage for an
octagonal patch with $264$ scatterers. Here, all edges have the same
length (which we choose to be $1$), wherefore the conductance and the
Hall voltage are periodic in the wave vector $q$ with period
$\pi$. However, the tile areas are incommensurate (the squares have an
area of $1$, the rhombs an area of $1/\sqrt{2}$), i.e., one has no
periodicity in the magnetic flux. Note that the result for the
Fibonacci grid is not periodic in either variable. Nevertheless, the
figures are reminiscent of the square lattice case.$^{6}$ In
particular, one clearly recognizes distinctive structures in the
transport properties of both systems, consisting of patterns that are
repeated quasiperiodically. A more detailed analysis of these features
will be presented elsewhere.

An experimental verification of our results could be based on a system
of interconnected small quantum wires in an appropriate setup at
sufficiently low temperatures to maintain phase coherence all over the
structure.

\end{document}